\documentclass[aps,prl,reprint,groupedaddress,nofootinbib]{revtex4-1}

\usepackage{amsfonts}
\usepackage{amsmath}
\usepackage{tikz}
\usetikzlibrary{arrows,calc}
\usetikzlibrary{patterns,snakes}

\newcommand{\bea}{\begin{eqnarray}}
\newcommand{\eea}{\end{eqnarray}}

\begin{document}

%\preprint{}

\title{Bound on Eigenstate Thermalization from Transport}


\author{Anatoly Dymarsky}
\affiliation{Department of Physics and Astronomy, \\ University of Kentucky, Lexington, KY 40506}
\affiliation{Skolkovo Institute of Science and Technology, \\ Skolkovo Innovation Center, Moscow, Russia, 143026}
%\thanks{}
%\altaffiliation{Skolkovo Institute of Science and Technology}
%\altaffiliation{University of Kentucky}


\date{\today}

\begin{abstract}
We show that macroscopic thermalization and  transport  impose constraints on matrix elements entering the Eigenstate Thermalization Hypothesis (ETH) ansatz and require them to be correlated. It is often assumed that the ETH reduces to Random Matrix Theory (RMT) below the Thouless energy scale. We show this conventional picture is not self-consistent. We prove that energy scale at which the RMT behavior emerges has to be parametrically smaller than the inverse timescale of the slowest thermalization mode coupled to the operator of interest. We argue that the timescale marking the onset of the RMT behavior is the same timescale at which  hydrodynamic description of transport breaks down.

\end{abstract}

%slow thermalization modes

%\keywords{}

\maketitle
Thermalization of isolated quantum systems has attracted significant attention recently. For the quantum ergodic systems without local integrals of motion
it is currently accepted that thermalization can be explained with the help of the Eigenstate Thermalization Hypothesis (ETH) \cite{jensen1985statistical,deutsch1991quantum,srednicki1994chaos,srednicki1996thermal,srednicki1999approach,
rigol2008thermalization,rigol2012alternatives,de2015necessity}. At the  technical level the ETH can be understood as an ansatz for the matrix elements of observables in the energy eigenbasis \cite{srednicki1999approach},
\bea
\label{ETH}
A_{ij}&=&A^{\rm eth}(E)\delta_{ij}+\Omega^{-1/2}(E) f(E,\omega)r_{ij},\\
E&=&(E_i+E_j)/2,\qquad \omega=E_i-E_j. \nonumber
\eea 
Here $A$ is an observable satisfying ETH \eqref{ETH}, $\Omega(E)dE$ is the density of states, $A^{\rm eth}$ and $f$ are smooth functions of their arguments, and $r_{ij}$ are pseudo-random fluctuations with unit variance.  The diagonal part of the ETH ansatz explains thermalization, at least in the sense that the expectation value of  $A$ in some initial state with mean energy $E$, after averaging over time, is equal to thermal expectation value of $A$ at the effective temperature $\beta^{-1}(E)=d\ln\Omega/dE$. The dynamics of thermalization is encoded in the off-diagonal matrix elements $r_{ij}$, as well as in the initial state $\Psi$, and is not universal.  In this paper we show that macroscopic thermalization, in particular  the type of transport present in the system, imposes constraints on the correlations of  $r_{ij}$.

Numerical studies confirm that the $r_{ij}$ behave ``randomly'' and oscillate around zero mean seemingly without any obvious pattern. Certainly the $r_{ij}$ can not be random in the literal sense as the form of $A_{ij}$ is fixed once the Hamiltonian and $A$ are specified. Moreover, $A$ often has to satisfy various algebraic relations. For example, in a spin lattice model one can choose $A$ to be a Pauli matrix acting on a particular site. In this case $A^2={\mathbb I}$, which requires $r_{ij}$ to be correlated. Similarly, the $r_{ij}$ can be constrained by the expected behavior of the four-point correlation function \cite{foini2018eigenstate,PhysRevLett.122.220601,PhysRevLett.123.260601,PhysRevE.104.034120}, etc. 

While the whole matrix $r_{ij}$ can not be random, there is a strong expectation that  fluctuations $r_{ij}$ can be treated as random if the indexes $i,j$ are restricted to belong to a sufficiently narrow energy interval. Assuming the interval  is centered around some $E$, we define $\Delta E_{\rm RMT}$ as the largest possible interval  such that all $r_{ij}$ with 
\bea
\label{box}
|E_i-E|, |E_j-E|\le \Delta E_{\rm RMT}/2,
\eea
can be treated for physical purposes as being random and independent (without necessary being normally distributed). %, without requiring $r_{ij}$ being normally distributed. 
The expectation that $r_{ij}$ reduces to a
Gaussian Random Matrix inside a sufficiently narrow interval is consistent with numerical studies which confirm that the $r_{ij}$ are normally  distributed \cite{beugeling2014finite,beugeling2015off,dymarsky2016subsystem} and that the form-factor $f$ becomes constant for $\omega$ smaller than inverse thermalization  timescale $2\pi/\tau$, called Thouless energy\footnote{Thouless energy $\Delta E_{\rm Th}$ is often defined as a scale of applicability of RMT to describe statistics of energy spectrum. Thermalization time $\tau$ is defined as time when the autocorrelation function
of an operator $A$  approximately saturates to a constant. The inverse scale $2\pi/\tau$ is the size of the ``plateau'' of $f^2(\omega)$, and is also called Thouless energy in the literature. For certain systems and operators probing slowest thermalization mode both quantities are known to coincide $\Delta E_{\rm Th}\approx 2\pi/\tau$ \cite{PhysRevLett.76.1130,d2016quantum,PhysRevB.104.L081112}.
}  %$\Delta E_{\rm Th}$
\cite{khatami2013fluctuation,d2016quantum,luitz2016anomalous,serbyn2017thouless}. Furthermore, for real symmetric $A_{ij}$ the variances of the diagonal and off-diagonal elements have been numerically shown to satisfy $\langle r_{ii}^2\rangle=2\langle r_{ij}^2\rangle$ \cite{dymarsky2017canonical,mondaini2017eigenstate,Richter:2020bkf}, which is consistent with and necessary for $r_{ij}$ to become a Gaussian Orthogonal Ensemble. Random behavior of $r_{ij}$ also naturally emerges in the recent attempt to justify ETH analytically \cite{anza2018eigenstate}. From the physical point of view the ``structureless''  form of $A_{ij}$ inside a small energy interval is expected on the grounds of  the hypothetical universal behavior of observables at late times \cite{cotler2017black,cotler2017chaos,santos2018nonequilibrium,PhysRevB.99.094312,PhysRevB.99.174313,Belin:2021ryy,Belin:2021ibv,Chen:2021qmm}. 

Reduction of $r_{ij}$ to an RMT  below $2\pi/\tau$ is seemingly in agreement with the 
conventional picture of thermalization. Assuming $\tau$ is the characteristic time of the slowest transport mode probed by $A$, after the time $t\gtrsim \tau$ 
the system will be in the ergodic regime, i.e.~value of $A$ will not be sensitive to the initial state. This suggests $r_{ij}$ should become structureless for $\Delta E_{\rm RMT}\sim 2\pi/\tau$ \cite{d2016quantum,serbyn2017thouless}. In this Letter we show this is not the case, and $\Delta E_{\rm RMT}$ has to be {\it parametrically} smaller than the Thouless energy $2\pi/\tau$.  

The key observation  is that the ETH ansatz \eqref{ETH} with random mutually-independent $r_{ij}$ is constrained by presence of states with extensively long thermalization times.  Let us consider an initial state $|\Psi\rangle$, which describes an out-of-equilibrium configuration with an order one overlap with the slowest mode probed by $A$. Then at late times 
\bea
\delta A(t,\Psi) \sim e^{-t/\tau},\qquad t\gtrsim \tau, \label{decay}
\eea
where
\bea
\label{deviation}
\delta A(t,\Psi)=\langle \Psi|A(t)|\Psi\rangle -\sum_i |C_i|^2 A^{\rm eth}(E_i).
\eea
Here second term is simply the equilibrated value of $A$, such that $\delta A$ asymptotes to zero at late times. We also assume  $|\Psi\rangle$  has  less than extensive energy variance $\Delta E$. While our argument is more general, 
for concreteness one can think of 
a 1D spin chain of length $L$ exhibiting diffusive transport of energy, and $A$ would be a local operator coupled to energy. 
In this case the initial state  can be taken to describe a quasi-classical configuration with an extensive displacement of energy, while  timescale in \eqref{decay} would be diffusive time $\tau\approx L^2/D$. 
An explicit construction of such a state $|\Psi\rangle$ is given in the Supplemental Materials (SM).  %A generalization for different types of transport is straightforward. 

%Using the bound on thermalization time $\tau\gtrsim L$ obtained  for a local $A$ above  and repeating the steps outlined below one can get $\Delta E_{\rm RMT}\sim o(1/L)$.
% If the system exhibits diffusive transport, provided $A$ is coupled to the diffusive quantity,  the time that the deviation \eqref{deviation} will remain of order one is even longer, $t \sim L^2$. In one dimensions the step-like profile discussed above can be decomposed into Fourier series with the $n$-th harmonics decaying as $e^{-n^2 D t/L^2}$,  where $D$ is the diffusion coefficient. At late times only the slowest mode survives with \eqref{deviation} behaving as $\sim e^{-t/\tau}$, $\tau=L^2/D$. This quasi-classical behavior for the state $|\Psi\rangle$ was recently confirmed numerically in  \cite{varma2017energy}. In what follows we will focus on the constraints provided by the diffusive modes. A generalization for different types of transport is straightforward. 

 To connect thermalization time $\tau$ to matrix elements of $A$ we introduce a parameter-dependent average, which is somewhat similar to the ``average distance'' used in \cite{garcia2017equilibration},  
\bea
\label{sin}
\langle \delta A\rangle_T\equiv \int_{-\infty}^{\infty} \delta A(t,\Psi) {\sin(2\pi t/T)\over \pi t}\, dt.
\eea
Here $T$ is a free parameter. When  $T$ becomes large, \eqref{sin}  reduced to the conventional average over time $T$.
After representing $A(t)$ in the energy eigenbasis using \eqref{ETH}  and performing the integral in \eqref{sin}  we find 
\bea
\label{sin2}
\langle \delta A\rangle_T&=&\langle \Psi|\delta A_T|\Psi\rangle,
\eea
where the operator $\delta A_T$ written in the energy eigenbasis has the form 
\bea
\label{dAT}
(\delta A_T)_{ij}=\left\{\begin{array}{cc}
\Omega^{-1/2}(E) f(E,\omega)r_{ij}, & |E_i-E_j|\leq 2\pi/T,\\
0 ,& |E_i-E_j|> 2\pi /T.
\end{array} \right. \, \,\, \, \,
\eea
In other words the matrix $(\delta A_T)_{ij}$ has a band structure, it coincides with $A_{ij}$ (after subtracting the non-random diagonal part) inside a diagonal band of size $2\pi/T$, and is zero outside.  This is schematically shown in Fig.~\ref{fig1}.
\begin{figure}[b]
\vspace{0.3cm}
\begin{tikzpicture}[scale=1]
    \coordinate (y) at (1,5);
    \draw[] (1,1) -- (1,6) --  (6,6) -- (6,1) -- (1,1);
    \draw[lightgray,fill] (1,5) -- (5,1) -- (6,1) -- (6,2) -- (2,6) -- (1,6)--(1,5) ;
    %\draw[] (2,6) -- (6,2);
    %
    %\draw[line width=0.5mm,dashed,blue] (1,4) -- (3,4) -- (3,6);
    %\draw[line width=0.5mm,dashed,blue] (2,5) -- (2,3) -- (4,3) -- (4,5) --(2,5);
    %\draw[line width=0.5mm,dashed,blue] (3,4) -- (3,2) -- (5,2) -- (5,4) -- (3,4); 	
    %\draw[line width=0.5mm,dashed,blue] (4,1) -- (4,3) -- (6,3); 	
    %
    %\draw[line width=0.5mm,blue] (3,4) -- (3,5) -- (2,5) --(2,4) --(3,4); 
    %\draw[line width=0.5mm,blue] (3,4) -- (3,3) -- (4,3) --(4,4) --(3,4); 	
    \draw[line width=0.5mm,dashed, black] (4,3) -- (4,2) -- (5,2) --(5,3) --(4,3); 	
\draw [
    thick,
    decoration={
        brace,
        mirror,
        raise=0.5cm
    },
    decorate
] (4.02,2.3) -- (5,2.3) 
node [pos=0.5,anchor=north,yshift=-0.55cm] {$2\pi/T$}; 	

\draw [
    thick,
    decoration={
        brace,
        mirror,
        raise=0.5cm
    },
    decorate
] (5.65,1) -- (5.65,6)
node [pos=0.5,anchor=east,xshift=1.5cm] {$\Delta E$}; 	 
\end{tikzpicture}
\caption{Visualization of the band matrix $(\delta A_T)_{ij}$ \eqref{dAT}.}
\label{fig1}
\end{figure}

The expectation value $\langle \Psi|\delta A_T|\Psi\rangle$  can be bounded by the largest  eigenvalue of  $\delta A_T$, which we denote by $x(T)$,
\bea
|\langle \Psi|\delta A_T|\Psi\rangle|\leq x(T). \label{L}
\eea  
Let us assume now that $T$ is sufficiently large such that $2\pi/T\leq \Delta E_{RMT}$. Then $(\delta A_T)_{ij}$  is a band random matrix with independent matrix elements and its largest eigenvalue is controlled by the variance function $\overline{(\delta A_T)^2_{ij}}=\Omega^{-1} f^2(\omega)$ \cite{molchanov1992limiting}. In the limit of a narrow band $T \Delta E \gg 1$, see SM, 
\bea
\label{bound}
x^2(T) = 8\int_0^{2\pi/T} f^2(E,\omega)\, d\omega.
\eea  
Technically,  \eqref{bound} assumes absence of correlations, while the definition of $\Delta E_{RMT}$  \eqref{box} does not exclude possible correlations of $r_{ij}$ and $r_{i'j'}$ along the diagonal, i.e.~when $(E_i+E_j)-(E_i'+E_j')$ is large while $|E_i-E_j|$ and $|E_{i'}-E_{j'}|$ are small. In SM we justify \eqref{bound} rigorously, using the result of \cite{dymarsky2017canonical},   by converting it into an inequality.  Looking ahead, our main result, inequality \eqref{inequality}, continue to hold with different numerical  coefficients.

%
%Then the sub-matrices $(A_T)_{ij}$   associated with $x(E', {\pi/(2T)},T)$ and $x(E'', {\pi/(4T)},T)$ are small enough to satisfy $\Delta E_{\rm RMT}\ge\Delta E$.  Then the corresponding sub-matrices are band random matrices with independent (but not necessarily normally distributed) $\delta A_{ij}$ and their respective largest eigenvalues are controlled by the variance function $\overline{\delta A^2_{ij}}(\omega)$ \cite{molchanov1992limiting}.  In particular largest eigenvalue satisfies the inequality \cite{dymarsky2017canonical} 

With help of \eqref{ETH} the integral  in the right-hand-side of \eqref{bound} can be expressed through the connected autocorrelation function of $A$ calculated at the effective inverse temperature $\beta^{-1}=d\ln\Omega/dE$ \cite{khatami2013fluctuation,luitz2016anomalous,d2016quantum},
\bea
\label{auto}
\langle A(t)A(0)\rangle_\beta \equiv \langle E|A(t)A(0)|E\rangle-\langle E|A(0)|E\rangle^2.
\eea
Now 
combining \eqref{L}  with \eqref{bound} written with help of \eqref{auto} we find the inequality, which should be satisfied so far $T\ge T_{\rm RMT}\equiv 2\pi/\Delta E_{\rm RMT}$,  
\begin{widetext}
\begin{align} \label{inequality}
|\langle \Psi|\delta A_T|\Psi\rangle|^2=\left|\int_{-\infty}^\infty \delta A(t,\Psi) {\sin(2\pi t/T)\over \pi t}dt \right|^2 \le  x^2(T)=
4\int_{-\infty}^\infty \langle A(t)A(0)\rangle_{\beta} {\sin(2\pi t /T)\over \pi t}dt\ ,
%\\  \left.\beta^{-1}=d\ln\Omega/dE\right|_{E_\Psi}=x^2(E,\Delta E,T).
\end{align}
\end{widetext}

The inequality \eqref{inequality} is our main technical result, which implies strong limitations on $\Delta E_{\rm RMT}$. As the characteristic size $L$  of the system grows, 
the autocorrelation function of $A$ approaches its thermodynamic form, which follows from quasi-classical hydrodynamic description, 
\bea
\langle A(t)A(0)\rangle_{\beta}\sim (t_D/t)^\alpha \label{2pt}
\eea
with some $L$-independent  $\alpha>0$ and $t_D$. Coefficient $\alpha$ depends on the type of transport $A$  couples to. The behavior  \eqref{2pt} applies for $t \gtrsim t_D$ and persists until $t\approx \tau$, after which 
the autocorrelation function becomes zero \cite{d2016quantum,luitz2016anomalous}.
Around the time $t\approx \tau$ the value of full autocorrelation function, i.e.~without the asymptotic value subtracted, should be inverse proportional to volume, indicating that the conserved quantity coupled to $A$ has spread across the whole system 
\bea
\left({t_D \over \tau}\right)^\alpha \propto {1\over L^d}. \label{tauL}
\eea
Here $L$ is a characteristic size of the system in dimensional units, e.g.~the number of spins, while $d$ is the number of spatial dimensions. 
Using \eqref{2pt}, and  for $T\gg t_D$ the right-hand-side of \eqref{inequality} can be approximated as follows, where we dropped all numerical coefficients, 
\begin{eqnarray}
\label{RHS}
\int_{0}^\infty \langle A(t)A(0)\rangle_{\beta} && {\sin(2\pi t /T)\over \pi t }dt \sim  \\  
& & \left\{ 
\begin{array}{l}
(t_D/T)^\alpha,\qquad\,    \tau \gtrsim T\gg t_D, \\
(t_D/\tau)^\alpha \tau /T,\ \  T\gtrsim \tau.
\end{array}  \nonumber
\right. 
\end{eqnarray}
For late times $T\gg t_D$ \eqref{RHS} is very small irrespective of the value of $\tau/T$.
Strictly speaking the estimate above is only correct far $\alpha < 1$ such that the integral  gets its main contribution for large $t$. In most cases this requires $d=1$.

The behavior of the left-hand-side of \eqref{inequality}  is quite different. Starting from  the expoential decay \eqref{decay} we find for large $T\gg \tau$,
\bea
\int_{0}^\infty \delta A(t,\Psi) {\sin(2\pi t/T)\over \pi t}dt \sim {\tau\over T}, \label{LHS}
\eea
which is in agreement with the qualitative picture that $\delta A(t,\Psi)$ remains of order one for the time $t\sim \tau$ and then quickly approaches zero. When $T$ is large but not necessarily larger than $\tau$ \eqref{LHS} remains of order one and the inequality \eqref{inequality} can not be satisfied. For \eqref{inequality} to be satisfied $T$ has to be parametrically larger than $\tau$, 
\bea
\left({\tau\over T}\right)^2\lesssim \left({t_D\over \tau}\right)^\alpha {\tau\over T}\quad  \Rightarrow\quad  T_{\rm RMT}\gtrsim   \tau\, L^d.  \label{MR}
\eea
To summarize, we see that the inequality \eqref{inequality} imposes a stringent bound on the energy scale $\Delta E_{\rm RMT}= 2\pi/T_{\rm RMT}$, which should be {\it parametrically} smaller than the Thouless energy $2\pi/\tau$. In particular, for a 1D diffusive system and a local operator $A$ coupled to conserved quantity we find 
\bea
T_{\rm RMT} \gtrsim  \tau L\sim  L^{3}\ .
\eea 
More generally, for any 1D system with local interactions transport can not be faster than ballistic,  $\tau \propto  L$,
and  therefore for any local operator,   
$T_{\rm RMT} \gtrsim \tau L\sim  L^2$.

\begin{figure}[b]
\includegraphics[width=0.49\textwidth]{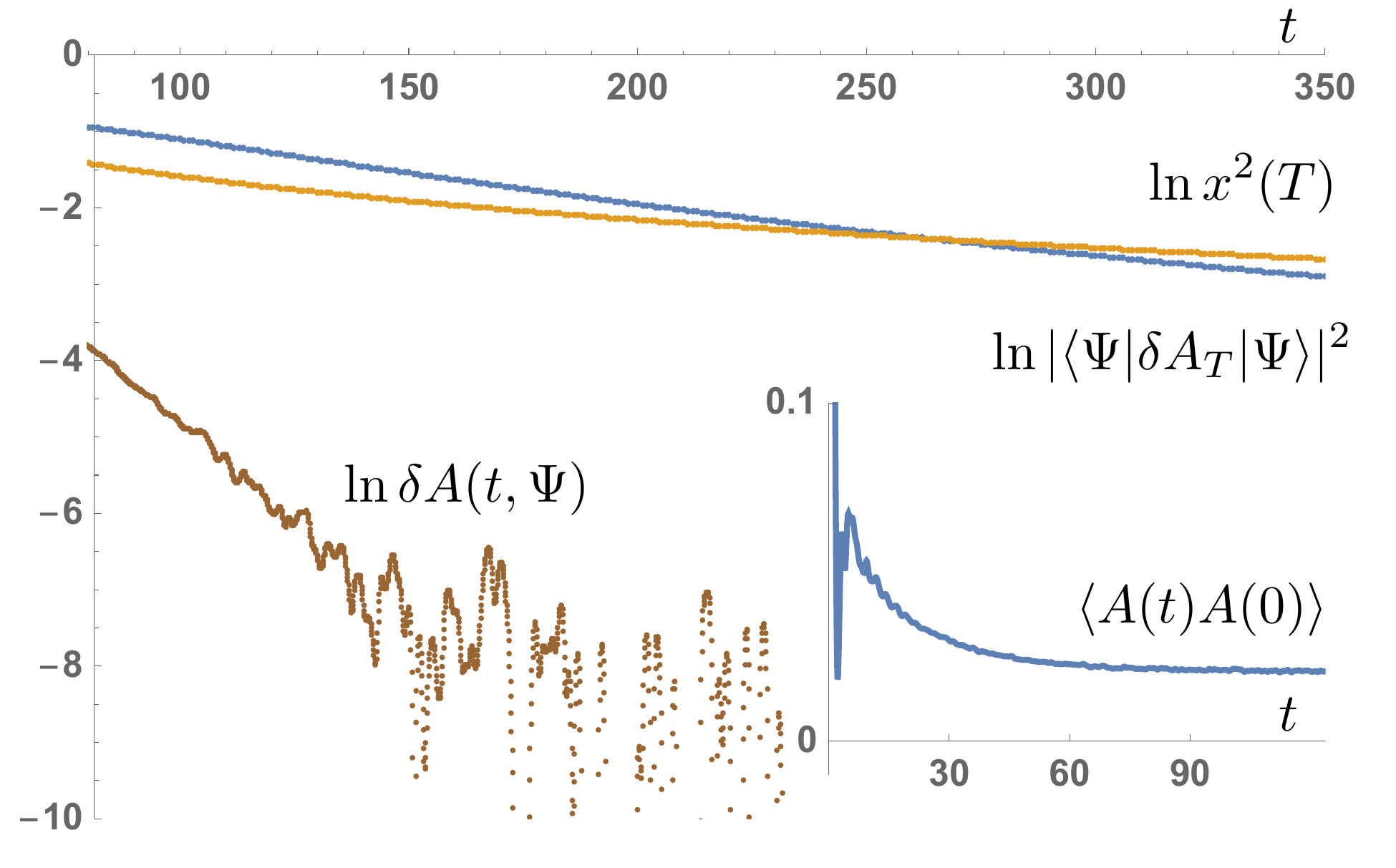}
\caption{Plots of the LHS and the RHS of \eqref{inequality} in logarithmic scale: $\ln |\langle \Psi|\delta A_T|\Psi\rangle|^2$ (blue) and $\ln x^2(T)$ (orange).  Also shown in brown $\ln \delta A(t,\Psi)$. Its approximately linear form (before saturation) confirms  exponential decay \eqref{decay}. Inset: plot of autocorrelation function. All plots are for non-integrable Ising spin chain with $L=24$ spins with open b.c., see SM for  details.}
\label{plotofinequality}
\end{figure}

We illustrate the inequality \eqref{inequality} and the resulting difference between $\Delta E_{\rm RMT}$ and $\tau^{-1}$ with help of an open non-integrable 1D Ising spin-chain with two polarizations of magnetic field.  The operator $A=\sigma_x^1$ is a one-site operator. 
This model is diffusive. In SM, where all technical details can be found, we numerically justify \eqref{decay} as well as \eqref{2pt} with $\alpha=1/2$.  The result,  the left-hand-side and the right-hand-side of  \eqref{inequality}, is shown in Fig.~\ref{plotofinequality}. The inequality is saturated for times $T$ significantly larger than thermalization time $\tau$, when the autocorrelation function plateaus (see the inset). 
This confirms the conclusion that the RMT time scale $T_{\rm RMT}$ is much larger than thermalization time. Smallness of $\tau/T_{\rm RMT}\ll 1$ was also recently confirmed numerically in \cite{Richter:2020bkf, Wang:2021mtp}.

For a translationally-invariant system it is also interesting to consider an operator $A_k$ with  a constant momentum.  Keeping in mind a 1D diffusive spin lattice system of length $L$, we denote by $A_{(m)}$ a local operator $A$ located at the site $m$. Then 
\bea
A_k={2^{1/2}\over L^{1/2}}\sum_{m=1}^L \cos\left(k\, m\right) A_{(m)},
\eea
where $L$ is dimensionless. 
The normalization factor $(2/L)^{1/2}$ is chosen such that the connected autocorrelation function is $L$-independent in the thermodynamic limit
\bea
\label{2ptk}
\langle A_k(t) A_{-k}\rangle_\beta\simeq e^{-t/\tau_k}\ ,\quad \tau_k \propto k^2/D.
\eea
With the same normalization the expectation value \eqref{deviation} in the  state
with a macroscopic amount of energy displaced
 will be 
\bea
\label{1pt}
\delta A(t,\Psi)\sim L^{1/2} e^{-t/\tau_k}
\eea
Although the $t$-dependence in \eqref{2ptk} and \eqref{1pt} is the same, different $L$-dependent prefactor will result in a constraint for $T_{\rm RMT}$. For large $T\gg \tau_k$ we can estimate 
\bea
\int_0^\infty {\sin(2\pi t/T)\over \pi t}e^{-t/\tau_k}  dt \sim {\tau_k\over T}
\eea
After ignoring unimportant numerical prefactors \eqref{inequality} yields, in agreement with \eqref{MR},
\bea
 T_{\rm RMT}\gtrsim \tau_k\, L.
\eea
%For fixed $k$, the initial configuration with the characteristic wavelength $1/k$ will require a finite $L$-independent time $1/(k^2 D)$ to  thermalize. Thus, we have shown that for  $A_k$  the inverse Random Matrix Theory scale  $\Delta E_{\rm RMT}^{-1}\gtrsim L/(k^2 D)$  is parametrically longer than the thermalization time.

To conclude, we have shown that the energy scale $\Delta E_{\rm RMT}$ at which the ETH ansatz reduces to Random Matrix Theory has to be parametrically smaller than the inverse thermalization time, i.e.~characteristic time of the slowest mode probed by the corresponding operator. For a 1D system and a local operator $A$ coupled to diffusive quantity we found $\Delta E_{\rm RMT}$ to be bounded by $(\tau L)^{-1}\sim L^{-3}$, where $L$ is the system size and $\tau\approx L^2/D$ is the diffusion time. 
%Our findings suggest that the conventional picture of thermalization of quantum ergodic systems, which assumes universal behavior of local observables at the scales below Thouless energy, is incomplete. In particular, there is an additional scale $\Delta E_{\rm RMT}$ relevant for thermalization dynamics.   

Our result (\ref{inequality},\ref{MR}) is an inequality, which raises the question of identifying the correct scaling of $\Delta E_{\rm RMT}$ with the system size and understanding significance of the associated timescale $T_{\rm RMT}=2\pi/\Delta E_{\rm RMT}^{-1}$ from the point of view of thermalization dynamics. We conjecture \eqref{MR} reflects the correct scaling $T_{\rm RMT}\propto \tau L^d$ and propose the following interpretation. The timescale $T_{\rm RMT}$ which marks the onset of random matrix behavior for an observable $A$ coincides with the end of macroscopic thermalization, i.e.~applicability of  hydrodynamic description of transport. The expectation value $\delta A(t,\Psi) \sim e^{-t/\tau}$ will decay exponentially until it saturates into exponentially small fluctuations of order $e^{-S/2}$, where $S \propto L^d$ is entropy. This happens around time
\bea
T \propto \tau S,
\eea 
which we conjecture to agree with $T_{\rm RMT}$ up to constant prefactors.
This interpretation, and scaling, is consistent with the onset of RMT-defined universal behavior of autocorrelation function at late times \cite{10.21468/SciPostPhys.9.3.034,PhysRevB.104.085117}. It is also consistent with the numerics shown in Fig.~\ref{plotofinequality}, where by the time the inequality \eqref{inequality} is satisfied the expectation value $\delta A(t,\Psi)$ has firmly saturated into the asymptotic fluctuation regime.

\begin{acknowledgments}
{\it Acknowledgments.}
I would like to thank Y.~Bar Lev, A.~Polkovnikov, and A.~Shapere for reading the manuscript. 
I also thank the University of Kentucky Center for Computational Sciences for computing time on the Lipscomb High Performance Computing Cluster.
I gratefully acknowledge support and hospitality of the Simons Center for Geometry and Physics, Stony Brook University at which part of the research for this paper was performed. This research is supported by the NSF under grant PHY-2013812.

\end{acknowledgments}

%\end{document}

%\clearpage

\section{Supplemental Material}
\subsection{Construction of initial  state}
Taking 1D lattice system (e.g.~spin chain) of length $L$ with open boundary conditions  as an example, in this section we explicitly construct an initial state $|\Psi\rangle$ with extensive thermalization time.  As  $A$ we take a local operator located at one of the edges of the system, for example one-spin operator acting on the first site.  

\begin{figure}
\includegraphics[width=0.4\textwidth]{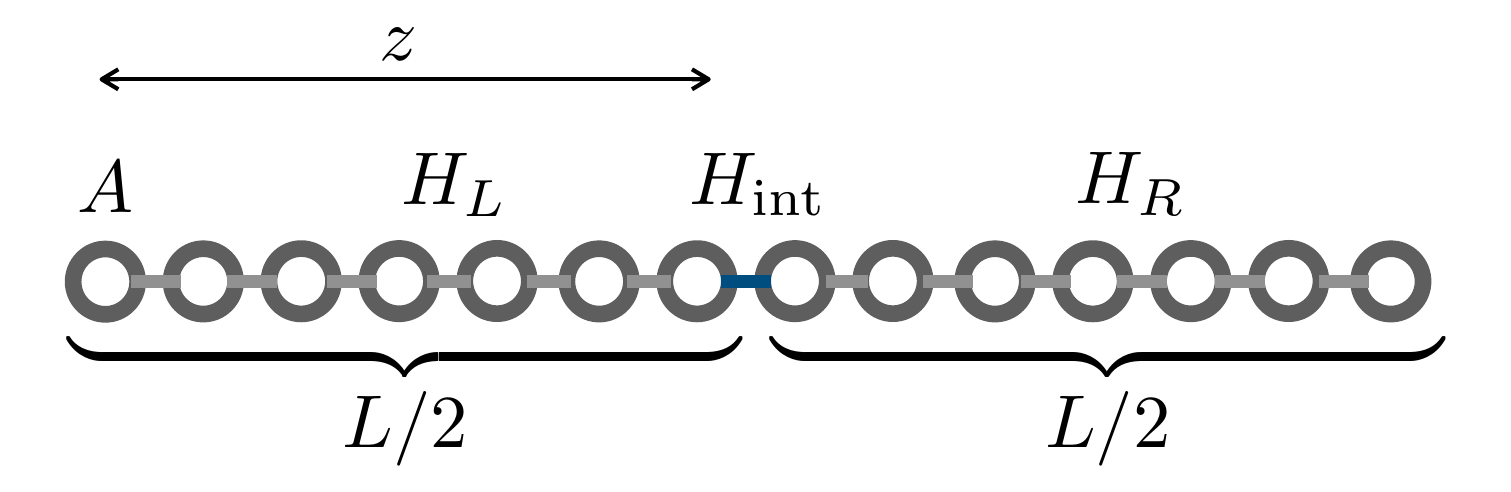}
\caption{Schematic visualization of the split \eqref{H0}. Circles correspond to lattice sites (spins). 
Interaction term $H_{\rm int}$ is visualized by the blue bond in the middle,  gray bonds correspond to $H_L$ and $H_R$. Distance between $A$ and $H_{\rm int}$ is denoted by $z$.} 
 \label{sc}
\end{figure}

We additionally assume the interactions are local, i.e.~the Hamiltonian $H$ only includes short-range interactions. 
Systems with local interactions have a finite maximal velocity of physical signals \cite{lieb1972finite,hastings2010locality}. As a result a quasi-classical configuration with an extensive amount of energy distributed locally would require at least time $t \gtrsim L$ to thermalize. To assign $|\Psi\rangle$ a particular effective temperature, we want  energy variance to be sub-extensive. Otherwise  the system may equilibrate, but not thermalize. To construct $|\Psi\rangle$ we split the system into two non-interacting subsystems of  approximately equal lengths $L_1\approx L_2$, $L_1+L_2=L$, by removing the corresponding interaction term(s) from the original Hamiltonian $H$,
\bea
H=H_L+H_R+H_{\rm int}\equiv H_0+H_{\rm int}. \label{H0}
\eea  
This split is schematically shown in Fig.~\ref{sc}.
The desired initial state can be chosen as a tensor product 
\bea
|\Psi\rangle =|E_L \rangle \otimes |E_R\rangle \label{initialstate}
\eea 
of two energy eigenstates of the corresponding subsystems $H_L,H_R$. By choosing different $E_L$ and $E_R$ with an extensive difference $|E_L-E_R|\sim L$, one creates a configuration of two adjacent subsystems with different effective temperatures.
Such states were previously studied numerically for diffusive systems in \cite{varma2017energy}.
 We will show now that in full generality it will take an extensive time for this state to thermalize. Indeed, $|\Psi\rangle$ is an eigenstate of the Hamiltonian $H_0$, which is the original Hamiltonian $H$ with the interactions between the two subsystems removed \eqref{H0}. Hence,  energy variance 
\bea
\nonumber
\Delta E^2=\langle \Psi|H^2|\Psi\rangle-\langle \Psi|H|\Psi\rangle^2=\\ \langle \Psi|H_{\rm int}^2|\Psi\rangle-\langle \Psi|H_{\rm int}|\Psi\rangle^2
\eea 
is bounded by the norm of $H_{\rm int}=H-H_0$ which is sub-extensive. In terms of the decomposition
\bea
\label{eigendecompositionSM}
|\Psi\rangle =\sum_i C_i |E_i\rangle\ ,
\eea 
where $|E_i\rangle$ are the eigenvalues of $H$, this means that most $|E_i\rangle$ contributing to \eqref{eigendecompositionSM} will correspond to the same energy density and therefore this state will thermalize rather than merely equilibrate  \cite{rigol2008thermalization}.  

To describe the time evolution of $|\Psi\rangle$ it is convenient to first switch to the Heisenberg picture and then employ the interaction picture splitting $H$ into $H_0+H_{\rm int}$. Then thermalization of $|\Psi\rangle$ is due to the growth of the local operator $H_{\rm int}$ under the time-evolution induced by $H_0$,
\bea
H_{\rm int}(t)=e^{i H_0 t}H_{\rm int}e^{-i H_0 t}.
\eea 
For a local operator $A$ located a distance $z$ away from the location of $H_{\rm int}$, the Lieb-Robinson bound \cite{lieb1972finite} guarantees that $\langle \Psi|A(t)|\Psi\rangle$ will remain constant to within an exponential precision at least up to times  $t\sim z$. This rigorously follows from the fact that $|\Psi\rangle$ is an eigenstate of $H_0$ and that, up to exponential corrections, $A$ and $H_{\rm int}(t)$ will commute up until times $t\sim z$.

Assuming the ETH \eqref{ETH} applies to $H_L$ and $H$, and $L$ is sufficiently large, we can estimate the  expectation value at $t=0$
\bea
\label{deviationSM}
\delta A(t=0,\Psi)=\langle \Psi|A|\Psi\rangle -\sum_i |C_i|^2 A^{\rm eth}(E_i)
\eea
to be $\delta A(t=0,\Psi)=A^{\rm eth}(E_L/L_1)-A^{\rm eth}(E/L)$, where $E\approx E_L+E_R$ is the total energy of $|\Psi\rangle$ and we changed the definition of $A^{\rm eth}$ to emphasize it is a smooth function of energy density.  Since energy densities $E/L$ and $E_L/L_1$ are different, \eqref{deviationSM} is non-zero, and will remain approximately the same for the period of time $t \gtrsim z\sim L_1 \propto L$, after which it will decay. In other words we have shown that the expectation value of a general local  operator $A$ in the state $\Psi$ will take an extensive time $\tau \gtrsim L$ to relax to its thermal value. (For local operators $A$ located near the middle point $z\ll L$, it is easy to construct a somewhat different initial $|\Psi\rangle$ reaching the same conclusion.) 

\begin{figure}
\includegraphics[width=0.45\textwidth]{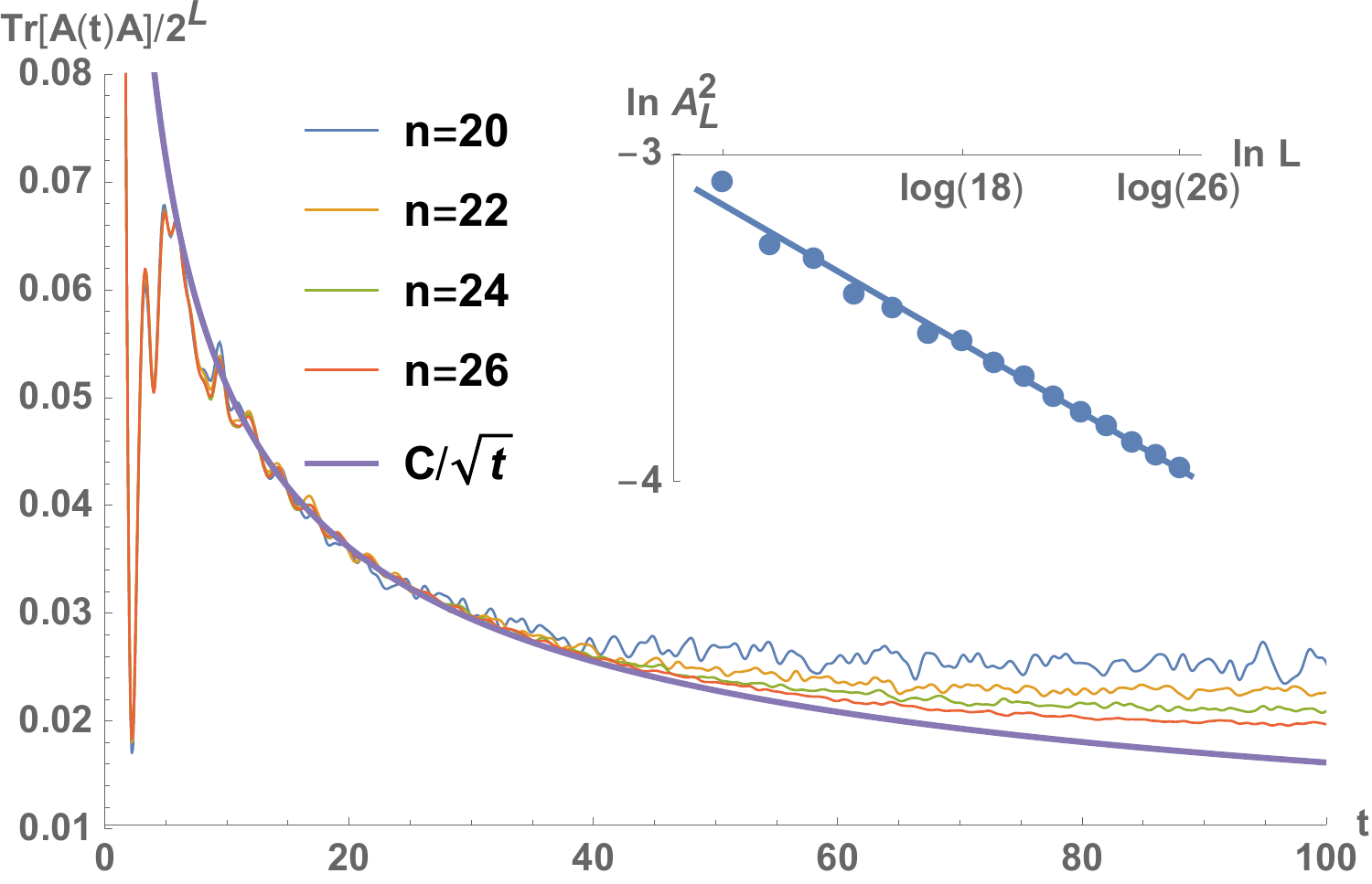}
\caption{Plot of $\langle A(t) A(0)\rangle_0^{\rm full}$ \eqref{infT} for different system sizes superimposed with $C/\sqrt{t}$ fit, where $C$ is a constant. Inset: asymptotic value  of \eqref{infT} $A_L^2$ vs $L$ plotted in log-log units, together with a linear fit.  The fitted slope $1.04$ is in a reasonable agreement with the asymptotic behavior \eqref{tauL} with $d=1$.}
\label{AA}
\end{figure}

The construction above is very general. While we have only proved, based on locality of interaction, that thermalization  of $\Psi$ will take linear in $L$ time, when the system is diffusive it will be quadratic in $L$. We demonstrate this numerically in the next section. 
\subsection{Numerical results}
\begin{figure}
\includegraphics[width=0.45\textwidth]{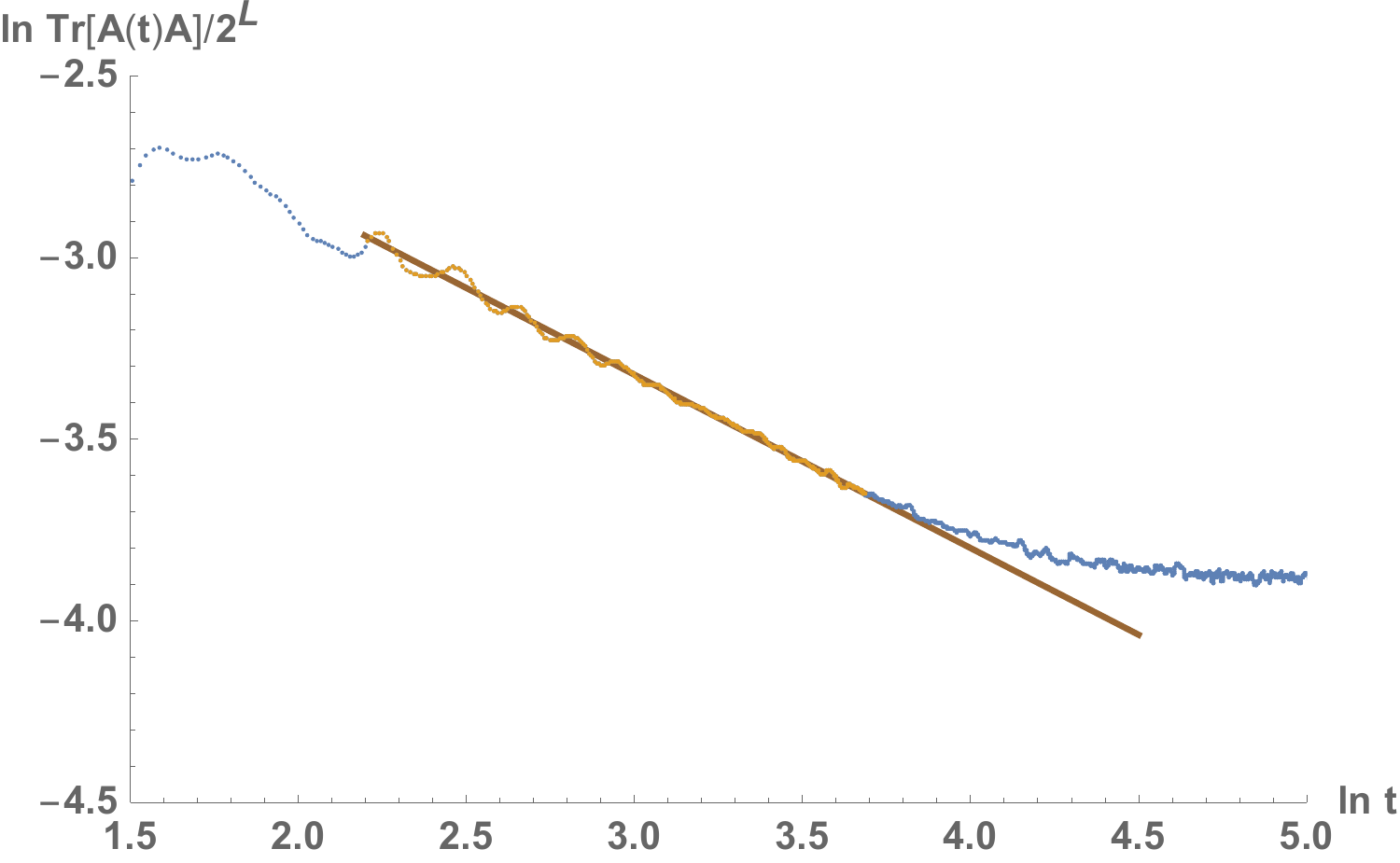}
\caption{Plot of $\langle A(t) A(0)\rangle_0^{\rm full}$ \eqref{infT} for $L=24$ together with the $(t_D/t)^\alpha$ fit plotted in log-log units. The fitted value of $\alpha=0.48$ is in good agreement with the diffusion value $\alpha=1/2$. Orange highlight marks the region used for the fit. It ends at the point taken as the Thouless time $\tau\approx 40$. Brown line is the linear fit (it extends beyond the orange region for visualization purposes).}
\label{AA24}
\end{figure}

We consider Ising spin-chain with two polarizations of magnetic field
\bea
H=\sum_{i=1}^{L-1} -\sigma_z^i \sigma_z^{i+1}+\sum_{i=1}^L (h\, \sigma_z^i+g\,\sigma_x^i), \label{Ising}
\eea
with $h=0.4$ and $g=1.05$. For these parameters the system satisfies ETH and exhibits diffusive transport, as we show below. As an observable we take $A=\sigma_x^1$. We first consider full autocorrelation function at infinity temperature
\bea
\label{infT}
\langle A(t) A(0)\rangle_0^{\rm full}={1\over 2^L}{\rm Tr}(A(t) A(0)).
\eea
Unlike \eqref{auto}, which is defined to asymptote to zero, \eqref{infT} will asymptote to a constant \eqref{tauL}. This is demonstrated in Fig.~\ref{AA} where we plot \eqref{infT}
for different system sizes $L\leq 26$. For numerics we use the typicality approach of \cite{PhysRevLett.110.070404} and therefore plots for smaller $L$ exhibit additional fluctuations. The functions are well fit by $C/\sqrt{t}$ with constant $C$ until they saturate into constant values $A^2_L$. The inset, showing $\ln(A_L^2)$ vs $\ln(L)$ confirms  $L$ dependence \eqref{tauL} with $d=1$. In Fig.~\ref{AA24} we plot \eqref{infT} for $L=24$ and fit it by  $(t_D/t)^\alpha$ with arbitrary $t_D$ and $\alpha$. The fit value $\alpha=0.48$ is reasonably close to diffusion value $\alpha=1/2$. This confirms the  behavior of autocorrelation function described in the main text. For $L=24$ Thouless time, defined as the approximate time of saturation of \eqref{infT}  is $\tau \approx 40$, see Fig.~\ref{AA24}.

To evaluate \eqref{inequality} we use the eigenstate version of the autocorrelation function $\langle A(t) A(0)\rangle_0$ \eqref{auto}, where the eigenstate $|E\rangle$ with $E\approx 0$ (infinite temperature) is evaluated numerically by  minimizing $H^2$ using the numerical approach of \cite{stathopoulos2010primme}. We checked that using $\langle A(t) A(0)\rangle_0$ or $\langle A(t) A(0)\rangle_0^{\rm full}$ after subtracting asymptotic value  lead to essentially equivalent results. 

\begin{figure}[b]
\includegraphics[width=0.45\textwidth]{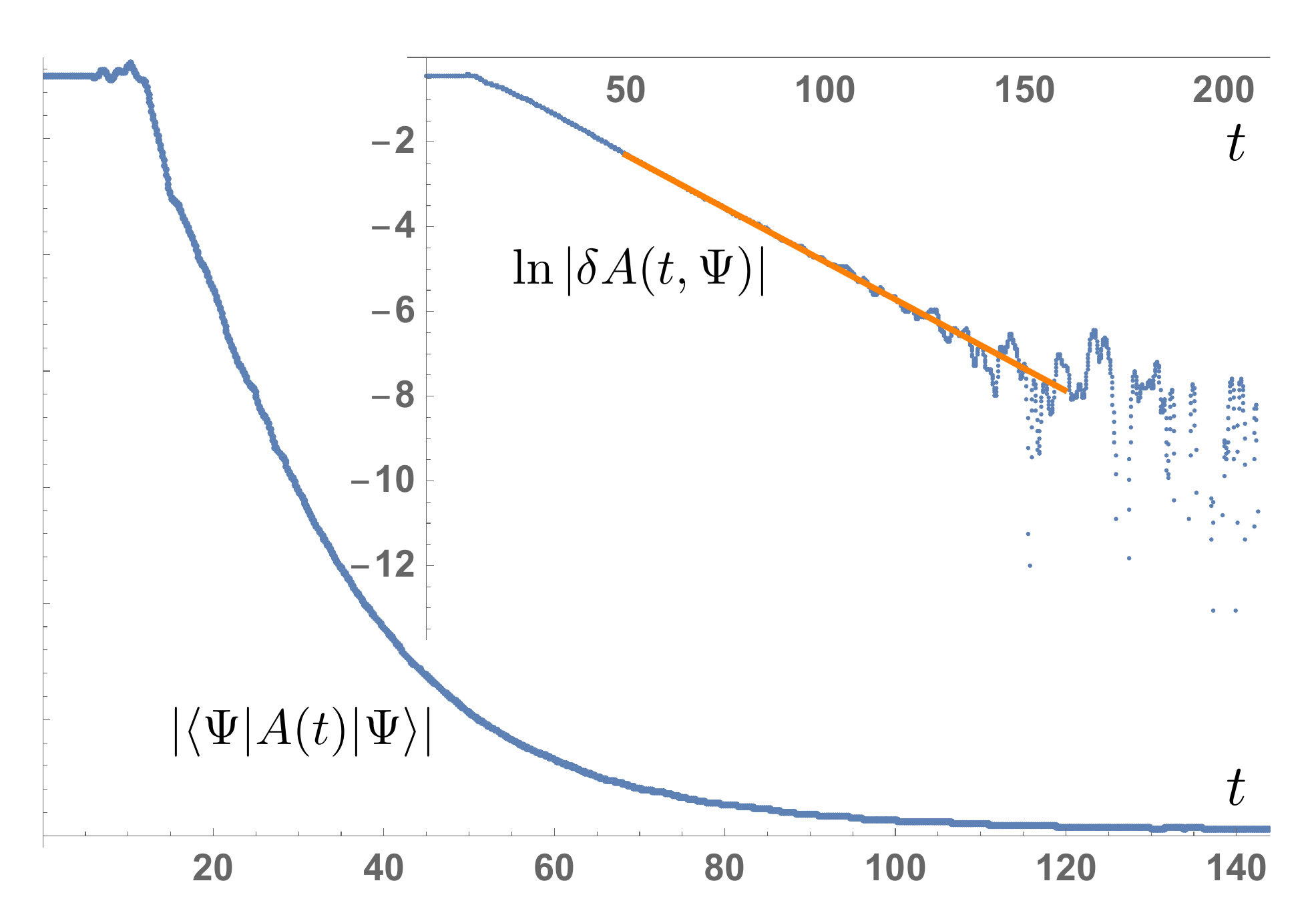}
\caption{Plot of $\langle \Psi  A(t) \Psi \rangle$ for the initial state  \eqref{is}.
Inset: the same plot (with the asymptotic value subtracted) in log scale, superimposed with a  linear fit. Very good quality of the linear fit before saturation confirms \eqref{decay}.}
\label{PAP}
\end{figure}

Next we turn to state $|\Psi\rangle$ which we construct using \eqref{initialstate} by combining two eigenstates for $L_1=11$ and $L_2=13$ subsystems,
\bea
\nonumber
|\Psi\rangle=|E_L\rangle \otimes |E_R\rangle,\,  E_L=-17.272283,\, E_R=16.920779. \\
\label{is}
\eea
This state has mean energy $E=-0.075627$ and variance $\Delta E=0.964163$.
The asymmetric division of $L_1\neq L_2$ is chosen to increase the value of $\delta A(t=0,\Psi)$ while keeping total energy of the state close to zero, which would correspond to infinite temperature. While the spectrum of Ising spin-chain \eqref{Ising} is approximately symmetric for moderate energies, there is a noticeable asymmetry between the ground state and the most excited state (ground state of $-H$). The ground state has largest value of $\langle E|A|E\rangle$, which is beneficial to make the inequality \eqref{inequality} stronger. Hence $|E_L\rangle$ is taken to be the ground state of $H_L$. 
To compensate the total energy to zero one has to take $|E_R\rangle$ to be the most excited state of $H_R$ but because of asymmetry, $L_2$ should be larger than $L_1$. Since both $L_1,L_2$ are sufficiently small, to find $|E_L\rangle, |E_R\rangle$ we use exact diagonalization, while time evolution was simulated using Chebyshev polynomials expansion. 
The result for $\langle \Psi|A(t)|\Psi\rangle$ is shown in Fig.~\ref{PAP}. With the subtracted asymptotic value, it is very well described by an exponential fit as shown in the inset, confirming \eqref{decay}. The value of diffusion time defined as the slope of $\ln \delta A(t)$ vs $t$ is equal $\tau\approx 19.82$. It  is significantly smaller than the value obtained above from the autocorrelation function. There is no contradiction here as both definitions have to reflect the same dependence on size and the diffusion constant $\tau \sim D L^2$ but can differ by the numerical prefactors.

\subsection{Maximal eigenvalue of band matrix}
\begin{figure}[b]
\begin{tikzpicture}[scale=1.1]
    \coordinate (y) at (1,5);
    \draw[] (1,1) -- (1,6) --  (6,6) -- (6,1) -- (1,1);
    \draw[lightgray,fill] (1,5) -- (5,1) -- (6,1) -- (6,2) -- (2,6) -- (1,6)--(1,5) ;
    %\draw[] (2,6) -- (6,2);
    %
    \draw[line width=0.5mm,dashed,blue] (1,4) -- (3,4) -- (3,6);
    \draw[line width=0.5mm,dashed,blue] (2,5) -- (2,3) -- (4,3) -- (4,5) --(2,5);
    \draw[line width=0.5mm,dashed,blue] (3,4) -- (3,2) -- (5,2) -- (5,4) -- (3,4); 	
    \draw[line width=0.5mm,dashed,blue] (4,1) -- (4,3) -- (6,3); 	
    \draw[line width=0.5mm,blue] (1,5) -- (2,5) -- (2,6) --(1,6) --(1,5); 
    \draw[line width=0.5mm,blue] (3,4) -- (3,5) -- (2,5) --(2,4) --(3,4); 
    \draw[line width=0.5mm,blue] (3,4) -- (3,3) -- (4,3) --(4,4) --(3,4); 	
    \draw[line width=0.5mm,blue] (4,3) -- (4,2) -- (5,2) --(5,3) --(4,3); 	
    \draw[line width=0.5mm,blue] (5,2) -- (5,1) -- (6,1) --(6,2) --(5,2); 	
    \draw[] (7.3,6)--(7,6)--(7,1)--(7.3,1);				
    \draw[] (7.7,6)--(8,6)--(8,1)--(7.7,1);
    \node[] at (7.5, 5.5)   (a) {\Large $\chi_1$};
    \node[] at (7.5, 4.5)   (a) {\Large $\chi_2$};
    \node[] at (7.5, 3.5)   (a) {\Large $\chi_3$};
    \node[] at (7.5, 2.5)   (a) {\Large $\chi_4$};
    \node[] at (7.5, 1.5)   (a) {\Large $\chi_5$};
    \draw[dashed] (7.2,2) -- (7.8,2);
    \draw[dashed] (7.2,3) -- (7.8,3);
    \draw[dashed] (7.2,4) -- (7.8,4);
    \draw[dashed] (7.2,5) -- (7.8,5);
\draw [
    thick,
    decoration={
        brace,
        mirror,
        raise=0.5cm
    },
    decorate
] (4,2.4) -- (5,2.4) 
node [pos=0.5,anchor=north,yshift=-0.55cm] {$2\pi/T$}; 	
\end{tikzpicture}
\caption{Schematic visualisation of a band matrix $\delta A_T$ and vector $|\chi\rangle$ represented as a sum of $\sum_{I=1}^N |\chi_I\rangle$ with $N=5$.}
\label{fig}
\end{figure}
A crucial step in the derivation of \eqref{inequality} is the expression \eqref{bound} for the largest eigenvalue of band matrix $\delta A_T$ \eqref{dAT}. If one assumes independence of $r_{ij}$ and also $\Delta E$ is small enough, such that the density of states is approximately constant $\beta \Delta E\ll 1$, matrix $(\delta A_T)_{ij}$ will become a band random matrix of the type studied in \cite{molchanov1992limiting},
\bea
(\delta A_T)_{ij}=r_{ij} {v(|i-j|/M)\over \sqrt{M}}, \label{RM}
\eea
where $r_{ij}$ are equally distributed independent random variables, $M\approx (\Omega \Delta E)\gg 1$ is the size of the matrix and $v$ is a smooth function of its argument. The resolvent of \eqref{RM}, which controls full density of states,  has to satisfy a particular integral equation. It can be solved in several special limits, when $v^2$ is a constant or when the band is infinitely thin, $v^2(t)=V^2 \delta(t)$. In these two cases maximal eigenvalue $x$ of $(\delta A_T)_{ij}$ are given by $x=2v$ and $=2V$
correspondingly. Translating to the notations of \eqref{dAT}
we find
\bea
V^2=\int_{-2\pi/T}^{2\pi/T} f^2(\omega) d\omega,
\eea
which yields \eqref{bound}.
More generally, largest eigenvalue of \eqref{RM} is bounded by $x^2\leq 4\int_{-1}^{1} v^2(t) dt$ \cite{dymarsky2017canonical}, a crucial result for what follows. Translating this into notations of \eqref{dAT} we find
\bea
x^2\leq 8\int_0^{2\pi/T}f^2(\omega)d\omega\equiv y^2. \label{b1}
\eea

Both assumptions, that $r_{ij}$ are mutually independent and $\beta \Delta E\ll 1$ are difficult to justify. In the case of the former, even if $T$ is sufficiently large such that $2\pi/T\leq \Delta E_{RMT}$, there still could be correlations between $r_{ij}$ and $r_{i'j'}$ along the diagonal, i.e.~when $(E_i+E_j)-(E_i'+E_j')$ is large but $|E_i-E_j|$ and $|E_{i'}-E_{j'}|$ are small. To rigorously justify \eqref{inequality}, instead of trying to evaluate the largest eigenvalue  of $\delta A_T$ we will obtain an upper bound in terms of function $f^2(\omega)$. The main idea is to split band  matrix $\delta A_T$ into many 
random square matrices of smaller size, over which we have better theoretical control, see Fig.~\ref{fig}. 

The square submatrices shown in Fig.~\ref{fig} have size $2\pi/T$ (solid blue) and $4\pi/T$ (dashed blue).
Assuming $T$ is sufficiently small, such that $4\pi/T\leq \Delta E_{RMT}$, due to  our main assumption outlined around \eqref{box} each of the square submatrices can be considered as random, with fully independent $r_{ij}$. If we further assume $4\pi\beta/T\ll 1$, within each square submatrix the density of states will be constant. Hence each of the square submatrices, both large (dashed lines) and small (solid lines), will be band random matrix of the type \eqref{RM} and their largest  by absolute value eigenvalues will be bounded by \eqref{b1}.

Now we consider band matrix $\delta A_T$ and split it into $N$ square submatrices of size $2\pi/T$ as is shown in Fig.~\ref{fig}. Since $\Delta E$ can be increased (by extending vector $|\Psi\rangle$ by zeros), we can take $N$ to be integer. 
Maximal eigenvalue of $\delta A_T$ can be defined via maximization problem 
\bea
\lambda(\delta A_T)=\max_\chi\, \langle \chi|\delta A_T|\chi\rangle
\eea
where maximization is over all normilzed $|\chi|^2=1$ states $|\chi\rangle =\sum_i c_i |E_i\rangle$ with $E_i\in [E-\Delta E/2,E+\Delta E/2]$ and otherwise arbitrary $c_i$.
We would like to introduce $N$ projectors $P_I$ associated with the small square submatrices, as is shown in Fig.~\ref{fig}, 
\bea
|\chi_I\rangle \equiv P_I |\chi\rangle,\quad |\chi\rangle=\sum_{I=1}^N |\chi_I\rangle,
\eea 
where $I=1\dots N$. We also introduce $|\chi_{I,I+1}\rangle \equiv |\chi_{I}\rangle+|\chi_{I+1}\rangle$ for $I=1\dots N-1$. The band structure of $\delta A_T$ ensures that 
$\langle \chi_I|\delta A_T|\chi_J\rangle=0$ unless $|I-J|\leq 1$. Therefore 
\bea
\nonumber
\langle \chi|\delta A_T|\chi\rangle=\sum_{I=1}^{N-1} \langle \chi_{I,I+1} |\delta A_T|\chi_{I,I+1}\rangle 
 -\sum_{I=2}^{N-1} \langle \chi_I |\delta A_T|\chi_{I}\rangle.
\eea 
Each matrix element of the form $\langle \chi_{I,I+1} |\delta A_T|\chi_{I,I+1}\rangle$ can be bounded by the largest by absolute value  eigenvalue of the large  (dashed line) submatrix, while each 
$ \langle \chi_I |\delta A_T|\chi_{I}\rangle$ can be bounded by the  largest by the absolute value eigenvalue of the small (solid line) submatrix. Since the largest eigenvalues of both large and small submatrices are bounded by $y$ \eqref{b1} we find 
\bea
\langle \chi|\delta A_T|\chi\rangle \leq  \left(\sum_{I=1}^{N-1} |\chi_{I,I+1}|^2   +
\sum_{I=2}^{N-1} |\chi_I|^2  \right)y. \label{in}
\eea 
After combining
\bea
\sum_{I=1}^{N-1} |\chi_{I,I+1}|^2=2-|\chi_1|^2-|\chi_N|^2\ ,\\
\sum_{I=2}^{N-1} |\chi_I|^2=1-|\chi_1|^2-|\chi_N|^2\ ,
\eea
with \eqref{in} we find the  generalization of \eqref{bound},
\bea
x^2(T)\leq 72\int_0^{2\pi/T} f^2(\omega)d\omega.
\eea
From here follows the inequality, which should be satisfied for $T\ge 2T_{\rm RMT}=4\pi/\Delta E_{\rm RMT}$, 
\begin{widetext}
\begin{align} \label{inequalitySM}
|\langle \Psi|\delta A_T|\Psi\rangle|^2=\left|\int_{-\infty}^\infty \delta A(t,\Psi) {\sin(2\pi t/T)\over \pi t}dt \right|^2 \le  x^2(T)\leq 
36\int_{-\infty}^\infty \langle A(t)A(0)\rangle_{\beta} {\sin(2\pi t /T)\over \pi t}dt\ .
%\\  \left.\beta^{-1}=d\ln\Omega/dE\right|_{E_\Psi}=x^2(E,\Delta E,T).
\end{align}
\end{widetext}

\bibliography{draft}

\end{document}